\documentclass[twocolumn,preprintnumbers]{revtex4}
\usepackage{amssymb}
\usepackage{amsfonts}
\usepackage{amsmath}
\usepackage{graphicx}
\usepackage{dcolumn}
\usepackage{bm}

\input{tcilatex}

\begin{document}

\title{ESR spectrometer with a loop-gap resonator for cw and time resolved
studies in a superconducting magnet}
\author{Ferenc Simon$^{\ast ,\dag }$, Ferenc Mur\'{a}nyi}
\affiliation{Budapest University of Technology and Economics, Physics and Solids in
Magnetic Fields Research Group of the Hungarian Academy of Sciences P. O.
Box 91, H-1521 Budapest, Hungary}
\date{\today}

\begin{abstract}
The design and performance of an electron spin resonance spectrometer
operating at 3 and 9 GHz microwave frequencies combined with a 9 T
superconducting magnet is described. The probehead contains a compact
two-loop, one gap resonator and is embedded in the variable temperature
insert of the magnet enabling measurements in the 0- 9 T magnetic field and
1.5-400 K\ temperature range. The spectrometer allows studies on systems
where resonance occurs at fields far above the $g$ $\approx 2$ paramagnetic
condition such as in strongly interacting spin systems. The low quality
factor of the resonator allows time resolved experiments such as e.g.
longitudinally detected ESR. We demonstrate the performance of the
spectrometer on the MgB$_{2}$ superconductor and the RbC$_{60}$ conducting
alkaline fulleride polymer.
\end{abstract}

\maketitle

\address{Budapest University of Technology and Economics, \\
Physics and\\
Solids in Magnetic Fields Research Group of the Hungarian Academy of Sciences\\
P. O. Box 91, H-1521 Budapest, Hungary, }


\begin{center}
{\large INTRODUCTION}
\end{center}

Recent developments in electron spin resonance instrumentation aims at the
use of high microwave frequencies and magnetic fields. This allows to obtain
higher resolution and the study of magnetic field and microwave frequency
dependent phenomena \cite{eatonreview,freedsum,hassan}. Most spectrometers
utilize a superconducting magnet with a resonant structure built in the bore
providing high sensitivity, such as Fabry-P\'{e}rot \cite%
{nijmegen,freedreson,smith}\ or cylindrical cavity resonators \cite%
{schmidt,hill}. An alternative is to use the superconducting magnet in
combination with a simple pass configuration that enables temperature
variability and multi-frequency operation at the cost of sensitivity \cite%
{nh3k3c60,mgb2}. Although, HF-ESR is seen to gain growing importance, the
conventional low-field ESR experiments, such as X-band, are usually required
to complement the spectroscopic information. It is often desired to perform
the low frequency experiments under similar experimental conditions than
those at the high frequencies e.g. at low temperatures (down to 1.5 K) that
is usually not available in commercial X-band spectrometers. A yet
unexplored domain of the frequency-field diagram is the use of low microwave
frequencies (\TEXTsymbol{<} 10 GHz) combined with high magnetic fields (6 T
or higher). This would enable to obtain extra information on strongly
correlated spin systems e.g. antiferromagnets. The primary obstacle for the
development of such spectrometers is the small usable size for a resonant
microwave structure required for measurements in superconducting magnet
bores. The recent developments of the NMR instrumentation beyond 1 GHz
enters the same domain with the same difficulties. A solution which
satisfies all these requirements would be an X-band spectrometer combined
with a superconducting magnet.

Here, we describe the development and performance of an ESR\ spectrometer
that allows the use of low frequencies in the S (2-4 GHz)\ and X (8-12 GHz)
bands and high magnetic fields (up to 9 T). A loop-gap resonator (LGR) with
coaxial leads is embedded in the variable temperature insert of a
superconducting magnet that is also used for high field/frequency ESR
measurements with a simple pass probehead. The LGR design is advantageous
when lower cavity quality and higher sample filling factors are required
such as for time resolved experiments for these microwave bands \cite%
{Froncisz,FronciszRSI,SchweigerRSI,MobiusRSI1997,MobiusRSI1998}. The
instrument has two orders of magnitude lower sensitivity than a commercial
ESP300 Bruker spectrometer that is compensated by the $\sim $20 times larger
usable samples providing comparable signal when sufficient sample amounts
are available. The apparatus has comparable sensitivity as the simple pass
HF-ESR spectrometer and allows studies on the same samples. The special
design of the LGR provides transparency for \textit{rf} radiation ($f$ 
\TEXTsymbol{<}10 MHz) and allows longitudinally detected ESR (LOD-ESR)
measurements at S and X bands. LOD-ESR complements continuous\ wave (cw) ESR
and spin-echo methods as it allows the measurement of short longitudinal
relaxation times \cite{pescia,strutz,atsarkin,MuranyiJMR} and to separate
overlapping ESR signals of spin species \cite%
{schweigerAMR,schweigerJMR,ablart,martinelli}. The instrument detects the
modulation of the longitudinal magnetization, $M_{z}$, induced by a chopped
microwave field, using a pick-up coil parallel to the external magnetic
field, $H_{0}$. LOD-ESR spectrometers were previously built at X-band and we
recently reported the development of the high field version \cite{MuranyiJMR}%
. The performance of the apparatus in the cw-ESR mode is demonstrated on the
MgB$_{2}$ superconductor below $T_{c}$ and in the LOD-ESR mode on the RbC$%
_{60}$ conducting alkaline fulleride polymer.


\begin{center}
{\large THE\ SPECTROMETER}
\end{center}

\begin{figure}[tbp]
\includegraphics[width=0.5\textwidth]{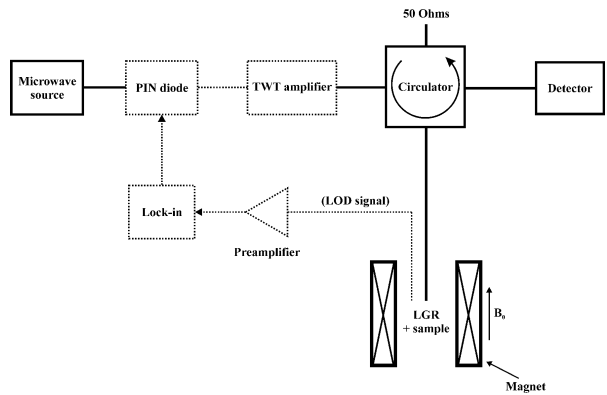}
\caption{Block diagram of the cw-ESR spectrometer with LGRs embedded in a
superconducting magnet. The optional elements for the LOD-ESR measurements
are shown in dashed boxes.}
\label{block_diagram}
\end{figure}

The block-diagram of the spectrometer operating in the cw-ESR mode including
the optional LOD-ESR elements (dashed) is shown in Figure \ref{block_diagram}%
. The microwave bridge is made of commercially available coaxial elements.
The coaxial microwave circuit is connected to the probehead with a flexible
coaxial cable (Pasternack Enterprises, PE3481) providing sufficient
mechanical stability. The source is an Agilent (Agilent 83751B) synthesized
frequency sweeper (2-20 GHz, maximal output power 20 dBm) that is locked to
the microwave resonator (see below). The microwaves are directed toward the
sample with a 4 port circulator (DITOM D4C2040 for 2-4 GHz and D4C8012 for
8-12 GHz) whose fourth port is terminated with a precision 50 Ohm
(Pasternack, PE6071). The microwave loss is 1 dB for the 2-4 GHz band and 3
dB at 8-12 GHz for the exciting and the reflected signals. The broad-band
microwave detector (Agilent 8474C, 0.01-33 GHz, sensitivity 500 mV/mW) does
not require microwave bias thus we do not employ a reference microwave arm.
Magnetic field modulation is driven by a lock-in (Stanford Research Systems
SRS830, 1 mHz-102 kHz) through a home-built U/I\ converter.

The instrumentation of the LOD-ESR experiments closely follows the previous
design of the HF-LOD-ESR apparatus \cite{MuranyiJMR} and here is only
outlined. A coaxial PIN diode (Advanced Technical Materials, S1517D, 0.5-18
GHz bandwidth, switching speed: 10-90 \%, 90-10 \%: 10 ns, insertion loss: 3
dB at 3 GHz and 5 dB at 9 GHz) provides the rapid microwave amplitude
modulation and at X-band a TWT microwave amplifier (Varian, TWTA VZX-6980GZ,
8-12.4 GHz, maximum output 40 dBm, amplification 40 dB) is also built in.
The PIN\ diode is driven by a high frequency lock-in (Stanford Research
Systems, SRS844, 0.025-200 MHz). An \textit{rf} parasite component with the
modulation frequency appearing on the output of the PIN diode is suppressed
out with a band-pass microwave filter consisting of two facing waveguide to
coaxial adapters (Agilent X281A). The LOD-ESR signal is detected with a
longitudinal coil (see also below)\ that is part of an \textit{rf}\ LC
circuit. The capacitors are placed outside the magnet to prevent temperature
drift of the \textit{rf} resonance frequency. The LOD signal is amplified
through a low noise \textit{rf} preamplifier (Analog Modules 322-6-50, 200
Hz-100 MHz, gain 40 dB, input noise 380 pV/$\sqrt{\text{Hz}}$) and is phase
sensitively detected by the SRS844 lock-in.

\begin{figure}[tbp]
\includegraphics[width=0.5\textwidth]{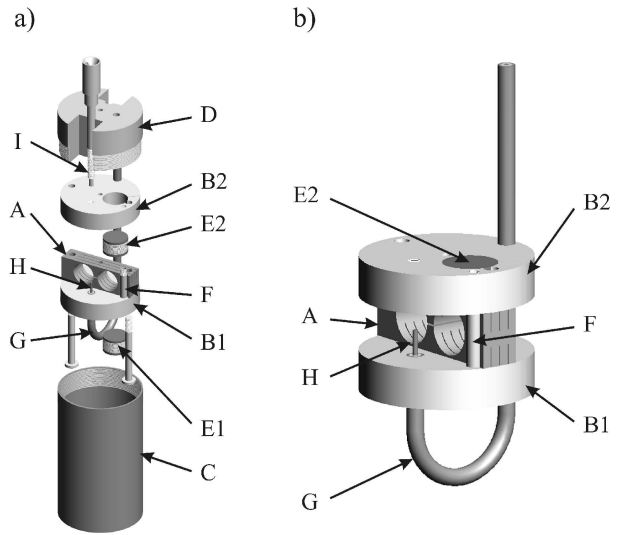}
\caption{a) Split and b) side views of the LGR probehead. A: LGR; B1, B2:
lower and upper microwave shields; C: microwave shielding cap; D: upper
support; E1, E2: modulation coils; F: tubing for modulation cable; G: curved
semirigid cable section; H: coupling antenna; I: coupling screw.}
\label{LGR_mount}
\end{figure}

The spectrometer is based on a high homogeneity (10 ppm/cm) superconducting
magnet (Oxford Instruments, 0-9 T) that is equipped with a variable
temperature insert (VTI) for the 1.5-400 K\ temperature range and is
regularly used for cw-HF-ESR measurements. Exchange gas in the sample
compartment (40 mm internal diameter) thermally connects it to the VTI.
Separation of the He gas in the sample compartment and the VTI avoids rapid
He pressure changes inside the resonator when the cooling rate is modified.
The outline of the probehead is shown in Figure \ref{LGR_mount}. Its outer
diameter is 38 mm. Except the LGR itself, all elements are made of brass
(CuZn39Pb3). The LGR (A) is placed between a lower (B1) and an upper (B2)
microwave shield which are surrounded by an on-screwable shielding cap (C).
These are fixed to the upper support (D) that is connected to the top of the
magnet with two stainless steel tubes (not shown). Heating resistance is
also attached to the upper support, the thermometer (Lakeshore, Cernox
CX1050-SD) fixed to the LGR provides precise temperature monitoring of the
sample. In cw-ESR mode, two coils (E1, E2) in an almost optimal Helmholtz
configuration provide modulation with 0.6 mT/A. The coils are connected with
a cable that is embedded in a brass tube (F) to prevent microphonics. The
upper coil is connected to a semi-rigid coaxial cable above the D support
with a short section of shielded flexible coaxial cable (not shown). We also
use a pair of coils to detect the LOD \textit{rf} signal in a
quasi-Helmholtz configuration (see below). The shielded cabling guarantees
shielding from the noisy environment, which is crucial for the LOD-ESR
studies \cite{MuranyiJMR}.

The microwaves are directed to the sample with a semi-rigid coaxial cable
(Pasternack, PE3931) that is connected to a curved coaxial cable section (G)
and ends in a capacitive antenna (H). Coupling is varied by a movable
grounded screw (I) that is uniaxial with the antenna following Ref. \cite%
{MobiusRSI1998}. The capacitive coupling is of smaller size and provides
enhanced mechanical stability as compared to the inductive one \cite%
{MobiusRSI1998}. The fine adjustment of the coupling is performed by moving
the coupling screw towards the antenna. On approaching the antenna, the
coupling increases. However, the coupling screw is only 3.8 mm apart from
the LGR and its movement affects the resonance frequency. To minimize the
required movement of the screw, the cavity is slightly undercoupled by
adjusting the effective length of the antenna before inserting the probehead
into the magnet.

The synthesized sweeper enables a simple design of the automatic frequency
control (AFC) as it allows the frequency control through a DC feed-back
signal and also AC modulation. Thus the AFC involves an analogue lock-in
(Stanford Research Systems SRS510)\ together with a simple circuit with
passive elements only.

\begin{figure}[tbp]
\includegraphics[width=0.5\textwidth]{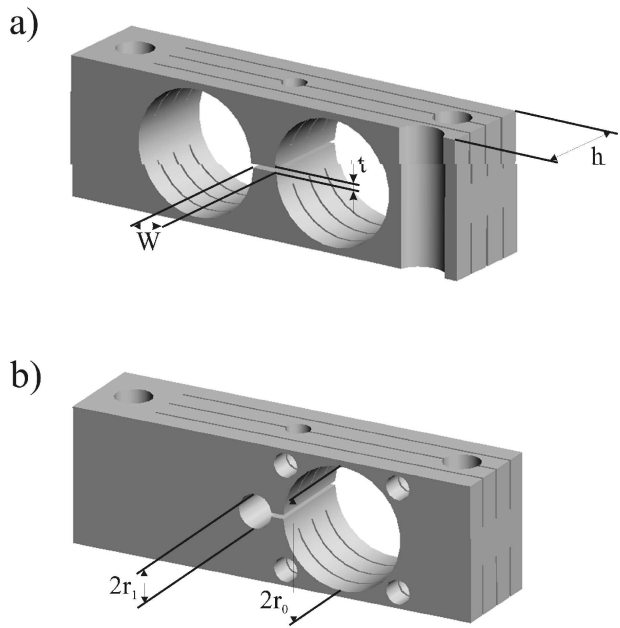}
\caption{a) The LGR resonator operating in S-band for cw-ESR measurements,
b) the LGR working at X-band, optimized for LOD-ESR measurements. }
\label{LGRs}
\end{figure}

The LGRs for the different experiments are shown in Fig. \ref{LGRs}. Fig. %
\ref{LGRs}a. and b. show the resonators used for cw-ESR measurements at 3
GHz and for LOD-ESR measurements at 9 GHz, respectively. Similar LGRs were
constructed for the 3 GHz LOD-ESR and 9 GHz cw-ESR (not shown). The
dimensions of the LGRs together with the calculated and measured resonance
frequencies and quality factors are summarized in Table I. The LGR\
parameter notations are taken from Ref. \cite{FronciszRSI} and are shown in
Fig. \ref{LGRs}. The calculated resonance frequency, $\nu _{1}$, is obtained
from the equivalent LC circuit using Eq. 2. in Ref. \cite{FronciszRSI} and
the resonant frequency with finite size corrections, $\nu _{2},$ is
calculated from Eq. 16 in Ref. \cite{FronciszRSI}. The small space available
for closing the flux inside the brass cap gives rise to a further $\sim $10
\% frequency up-shift. The design allows the use of cylindrically shaped
samples with diameters up to 8 mm and a height of 5 to 8 mm that are used
for the simple pass HF-ESR apparatus with 8 mm stainless steel tubing, such
as a loose powder contained in a Teflon sample holder of a metallic sample
as RbC$_{60}$ \cite{JanossyPRL} or fine grains of MgB$_{2}$ cast in epoxy 
\cite{mgb2}.

\begin{table}[tbp]
\caption{Dimensions, resonant frequencies and quality factors of the LGRs in
this work. $r_{0}$ and $r_{1}$ are the radii of the sample and flux return
loops, respectively, $W$ is the width, $t$ is the separation of the slit, $h$
is the height of the LGR (all length parameters are in mm, frequencies in
GHz units). Q is the quality factor of the critically coupled unloaded
cavities.}%
\begin{tabular}{lllllllllll}
\hline\hline
LGR & $r_{0}$ & $r_{1}$ & $W$ & $t$ & $h$ & $\nu _{1}$ & $\nu _{2}$ & $\nu
_{EXP}$ & Q(copper) & Q(brass) \\ \hline
3 GHz & 4.5 & 4.5 & 2 & 0.3 & 8 & 3.26 & 3.25 & 3.55 & 500 & - \\ 
9 GHz & 4.5 & 1.5 & 1 & 0.3 & 8 & 8.41 & 8.33 & 8.99 & 350 & 250 \\ 
\hline\hline
\end{tabular}%
\end{table}
\label{dimensions}

The thin upper and lower walls of the LGR and the cuts perpendicular to the
axis of the loops minimize eddy currents in the field modulation detected
cw-ESR experiments. For LOD-ESR measurements, we use cavities made of the
lower conducting brass to obtain better transparency for the detected 
\textit{rf} signal. The pick-up coil is wound through the holes indicated in
Fig. \ref{LGRs}b. in a quasi-Helmholtz configuration providing good
transparency for the \textit{rf} signal. The interaction between microwaves
and the pick-up coil does not affect the pick-up signal. In the LOD-ESR
mode, we have 12 dBm at 9 GHz and 16 dBm at 3 GHz at the sample. At 9 GHz,
we can use a TWT amplifier (see Fig. \ref{block_diagram}) for the LOD-ESR
measurements. The 4 port circulator limits the maximum power to 36 dBm peak
or 33 dBm cw power. At low temperatures the power is limited to lower values.

\begin{center}
{\large PERFORMANCE OF THE SPECTROMETER}
\end{center}

\begin{figure}[tbp]
\includegraphics[width=0.4\textwidth]{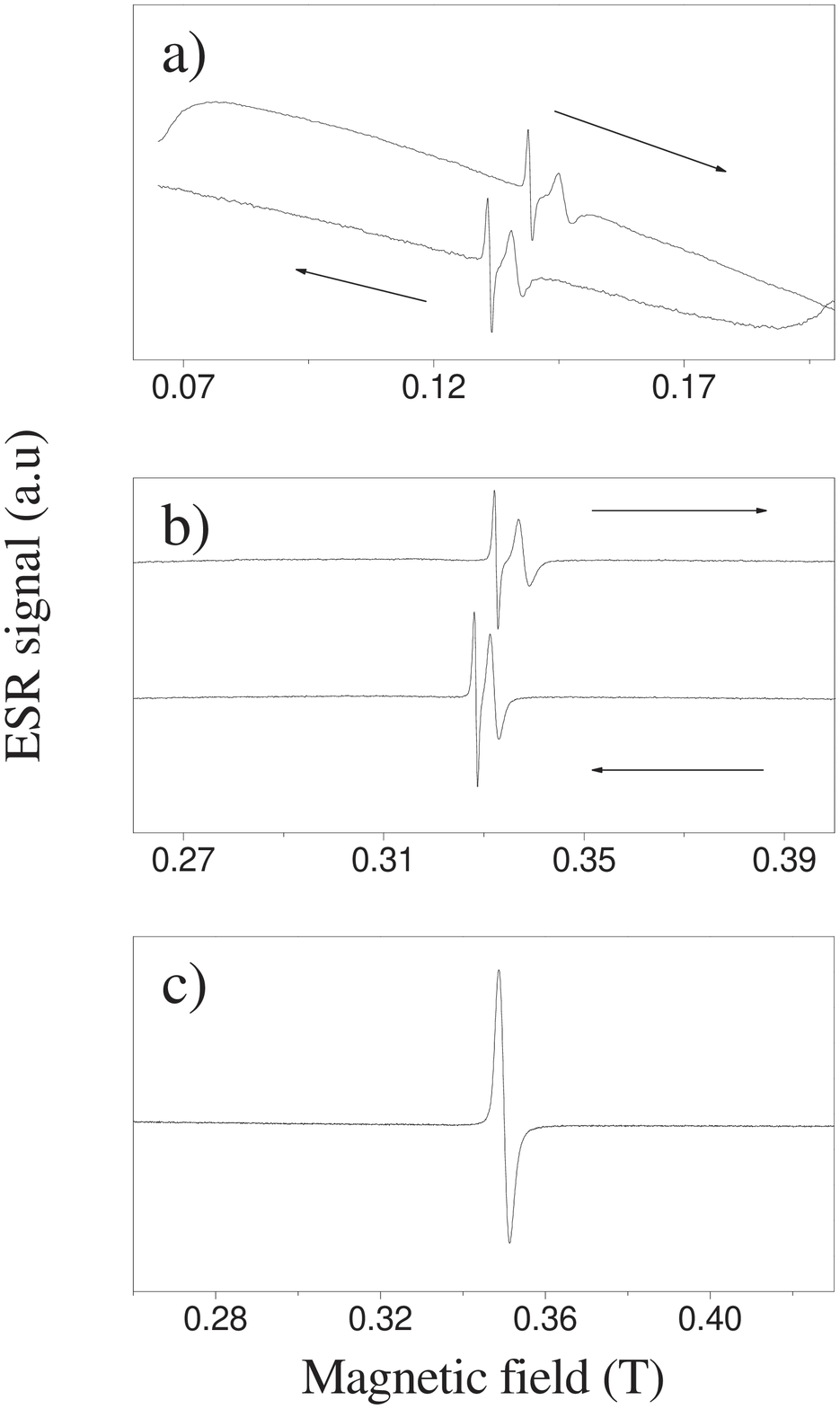}
\caption{MgB$_{2}$ spectra measured with the different spectrometers at 30
K. a) Measured in the S-band LGR at 3.71 GHz, b) measured in the X-band LGR
at 9.14 GHz, c) measured in a Bruker ESP300 commercial spectrometer at 9.38
GHz. The narrow line at a) and b) is a KC$_{60}$ for the field calibration,
arrows indicate the field sweep direction. Note the significant hysteresis
loop and the non-linear background in the S-band measurement due to the
superdconductvity in the sample.}
\label{cw_spectra}
\end{figure}

\begin{center}
The cw-ESR spectrometer
\end{center}

In Fig. \ref{cw_spectra}, we show typical cw-ESR spectra of MgB$_{2}$ in the
superconducting state of this material. MgB$_{2}$ is a highly conducting
material in its normal state and becomes a superconductor below $T_{c}$= 39
K \cite{akimitsu}. The spectra were measured by the currently described
spectrometer (a and b) at S and X-band, respectively, as well as on a Bruker
ESP300 commercial spectrometer at X-band (c). In the fine powder of the MgB$%
_{2}$ superconductor the conducting particles were separated by a powder of
the non-magnetic SnO$_{2}$ insulator \cite{mgb2,Simoncondmat}. Measurement
in the superconducting magnet requires a magnetic field calibrating $g$%
-factor standard. We used KC$_{60}$ that has a narrow line with a
temperature independent intensity and $g$-factor. Spectra of comparable
quality can be obtained with the current LGR\ and the commercial
spectrometers. The calculated sensitivity of the current spectrometer is $S=$%
2$\cdot $10$^{13}$ spin/G and $S=$5$\cdot $10$^{12}$ spin/G at for the S and
X band LGRs, respectively. These values have to be compared to the $S=$5$%
\cdot $10$^{10}$ spin/G of the ESP300 spectrometer determined on the same
samples. The lower sensitivity can be partially compensated by the
possibility of using up to 20 times larger amounts of the same sample
without affecting the resonance mode of the cavity. As a result, we found
that the current spectrometer provides only a factor 3-4 times smaller
signal to noise ratios than the commercial apparatus when there is no limit
on the sample availability. The sensitivity of the LGRs is comparable to
that of the simple pass HF-ESR apparatus $S=2\cdot $10$^{13}$ spin/G at 75
GHz \cite{Janossyunpublished}.

An advantage of the lower $Q$ of the LGR is that it allows the detection of
ESR\ in superconductors in magnetic fields as low as $\sim $ 0.1 T as Fig. %
\ref{cw_spectra}a demonstrates. The so-called vortex noise has been a
limiting factor of field modulated ESR studies in superconductors using high 
$Q$ commercial cavities that could be overcome only by avoiding magnetic
field modulation \cite{JanossyPhysC}. Detailed temperature dependent study
of MgB$_{2}$ have been published separately \cite{Simoncondmat}.

\begin{figure}[tbp]
\includegraphics[width=0.4\textwidth]{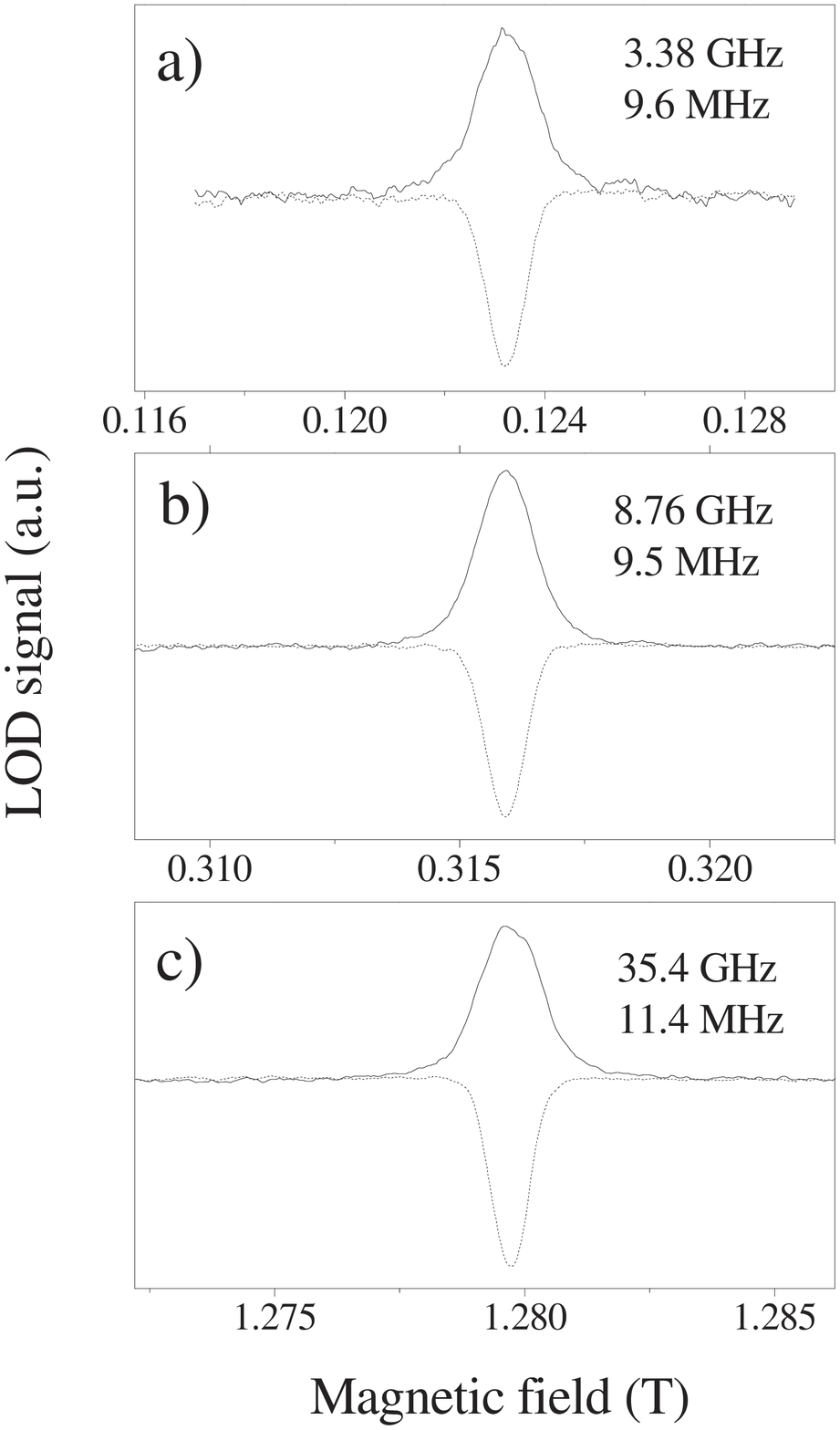}
\caption{Comparison of the LOD-ESR performance at several microwave
frequencies, a) S-band LGR, b) X-band LGR, c.) Q-band simple pass 
\protect\cite{MuranyiJMR} (solid and dashed lines are in and out-of-phase
LOD signals, respectively). Insets show the microwave and the \textit{{rf}
modulation frequencies.}}
\label{LOD_spectra}
\end{figure}

\begin{center}
The LOD-ESR spectrometer
\end{center}

In Fig. \ref{LOD_spectra}, we compare the performance of our spectrometer
operating in the LOD mode with the previously developed 35 GHz simple pass
LOD-ESR apparatus \cite{MuranyiJMR}. We used RbC$_{60}$ to calibrate and
test the performance of our system. The LOD-ESR signal is proportional to
the ESR saturation factor, $\gamma _{e}^{2}B_{1}^{2}T_{1}T_{2}$, where $%
\gamma _{e}$, $B_{1}$, $T_{1}$ and $T_{2}$ are the electron gyromagnetic
factor, exciting microwave field, longitudinal and transversal relaxation
times, respectively. Compared to the simple pass ($Q=1$)\ 35 and 75 GHz
LOD-ESR spectrometers, a $Q$-fold increase in the signal is expected in the
current spectrometer. The use of smaller magnetic fields reduces the
longitudinal magnetization and the corresponding LOD signal. Altogether, one
order of magnitude better performance is expected at the same power than
with the 35 GHz spectrometer. However, we observed experimentally that the
signal amplitudes are similar. The probable origin of the 10 times smaller
sensitivity than expected is that the coil and microwave filling factors are
smaller than for the 35 GHz LOD-ESR spectrometer. The quasi-Helmholtz
configuration for the pick-up coils is known to have \cite{HoultJMR} a
smaller filling factor than the solenoid used at 35 GHz \cite{MuranyiJMR}.
The magnetic component of the microwave field inside the cavity is highly
inhomogeneous and a substantial part of the sample is not in the maximum
field position. This yields an experimentally determined $S=$7$\cdot $10$%
^{16}$ spin/G$^{3/2}$ at 3.5 GHz (16 dBm) and $S=$1.5$\cdot $10$^{16}$ spin/G%
$^{3/2}$ at 9 GHz (at 12 dBm), this has to be compared to the 35 GHz simple
pass LOD ESR that had $S=$3.3$\cdot $10$^{16}$ spin/G$^{3/2}$ at 12 dBm \cite%
{MuranyiJMR}. At 9 GHz the TWT power amplifier at 33 dBm output power allows
to achieve a maximum sensitivity of $S=$1.5$\cdot $10$^{14}$ spin/G$^{3/2}$.

RbC$_{60}$ has relaxation times in the 13-46 ns range that cannot be
measured with usual spin-echo methods. On the other hand, LOD-ESR has been
successful in determining $T_{1}$ in this compound \cite{atsarkin,MuranyiJMR}%
. The full method of obtaining $T_{1}$ data from the raw LOD-ESR\ data was
reported previously \cite{MuranyiJMR} and is only outlined here. In the
simplest case \cite{pescia,atsarkin} when $\Omega \cdot T_{1}<1$, the ratio
of the in ($u$) and out-of phase ($v$) components of the oscillating
longitudinal magnetization gives: $T_{1}=\frac{1}{\Omega }v/u$. The accuracy
of measuring $T_{1}$ relies not only on the magnitude of the detected
LOD-ESR signal but also on the microwave amplitude modulation frequency,
this latter being optimal when $\Omega \approx 1/T_{1}$. The maximum
modulation frequency of our apparatus is limited to 10 MHz by the bandwidth
of the LGRs used and the switching speed of the PIN\ diode. The best
accuracy is for $T_{1}$ of 16 ns. However, $T_{1}$'s down to 5 ns can still
be measured with reduced accuracy.


\begin{center}
{\large CONCLUSION}
\end{center}

In conclusion, we presented the construction and the performance of a
spectrometer that utilizes low microwave frequencies and is embedded in the
VTI of a superconducting magnet. The spectrometer is made of commercially
available elements and its performance is comparable to the commercially
available resistive magnet based S and X-band spectrometers. Thus, our
system is an affordable alternative to the commercial low-frequency
spectrometers and it can be readily installed at an HF-ESR laboratory. In
addition, the system allows time resolved ESR experiments such as the
detection of longitudinally detected ESR thus allowing the measurement of
very short $T_{1}$'s in metallic samples. The current spectrometer completes
the previously developed 35 GHz version of LOD-ESR enabling magnetic field
dependent relaxation studies. We plan to perform experiments on systems
where combination of low frequencies and high magnetic fields is required
such as in antiferromagnets \cite{nanio2}. The approach reported here may
also find applications in the emerging ultra-high field NMR applications as
customary NMR frequencies approach the 1 GHz value \cite{clark} where our
resonator design, completed with frequency tunability appears to be a
feasible design.


\begin{center}
{\large ACKNOWLEDGEMENTS}
\end{center}

This work is dedicated to the memory of L\'{a}szl\'{o} Berende. The authors
would like to express their gratitude to A. J\'{a}nossy for helping the
development of the current spectrometer and for many useful discussions. L.
Forr\'{o} is acknowledged for providing the RbC$_{60}$ sample and for
allowing the use of the ESP300 spectrometer. C. Petrovic, S. Bud'ko, and P.
Canfield are acknowledged for the MgB$_{2}$ sample. A. Sienkiewicz is
gratefully acknowledged for suggesting the design of the resonator. B. Horv%
\'{a}th is acknowledged for the technical assistance. Support from the
Hungarian State Grants, OTKA T043255, OTKA TS040878, OTKA NDF45172 and FKFP
0352/1997 are acknowledged. FS acknowledges the Bolyai Hungarian Research
Fellowship and the PATONN Marie-Curie MEIF-CT-2003-501099 grants for support
and the hospitality of the University of Vienna during the preparation of
the manuscript.

$^{\ast }$corresponding author:\ simon@esr1.fkf.bme.hu

$^{\dag }$Current address: Institut f\"{u}r Materialphysik, Universit\"{a}t
Wien, Strudlhofgasse 4, A-1090, Wien, Austria


\bigskip

\end{document}